\documentclass[prl,twocolumn,showpacs,amsmath,amssymb,superscriptaddress]
{revtex4}

\usepackage[dvips]{color}
\usepackage{graphicx}

\begin{document}

\title{Phase coherence, visibility, and the superfluid--Mott-insulator 
transition on one-dimensional optical lattices}
  
\author{P. Sengupta}
\affiliation{Department of Physics, University of California, 
Riverside, CA 92521, USA}
\author{M. Rigol}
\affiliation{Physics Department, University of California, Davis,
CA 95616, USA}
\author{G. G. Batrouni}
\affiliation{Institut Non-Lin\'eaire de Nice, UMR 6618 CNRS,
Universit\'e de Nice--Sophia Antipolis, 1361 route des Lucioles,
06560 Valbonne, France}
\author{P. J. H. Denteneer}
\affiliation{Lorentz Institute, Leiden University, P. O. Box 9506,
2300 RA  Leiden, The Netherlands}
\author{R. T. Scalettar}
\affiliation{Physics Department, University of California, Davis,
CA 95616, USA}

\begin{abstract}
We study the phase coherence and visibility of trapped atomic condensates on 
one-dimensional optical lattices, by means of quantum Monte-Carlo 
simulations. We obtain structures in the 
visibility similar to the kinks recently observed experimentally 
by Gerbier {\it et al.}~[Phy. Rev. Lett.  {\bf 95}, 050404 (2005); 
Phys. Rev. A {\bf 72}, 053606 (2005)]. We examine these features 
in detail and offer a connection to the evolution of 
the density profiles as the depth of the lattice is 
increased. Our simulations reveal that as the interaction strength, 
$U$, is increased, the evolution of superfluid and 
Mott-insulating domains stall for finite intervals of $U$.
The density profiles do not change with 
increasing $U$.  We show here that in one dimension the visibility 
provides unequivocal signatures of the melting of Mott domains with 
densities larger than one.
\end{abstract}


\maketitle

The realization of trapped Bose-Einstein condensates in ultracold
atoms on 
optical lattices has opened up the possibility of observing experimentally 
various quantum phases -- e.g., superfluid (SF) and Mott-insulator
 (MI) --  and the study of the nature of the transitions between them in a 
well-controlled manner. Indeed, the existence of SF and MI 
phases on optical lattices was established experimentally 
\cite{greiner,StofEss04}, where it was demonstrated that by increasing 
the optical lattice depth  
the system passes from a SF phase to a predominantly MI one. Contrary 
to the unconfined case, in traps there is in general a coexistence of 
SF and MI domains. Hence, the passage from SF to 
MI has to be understood as a cross-over rather than as a quantum phase 
transition \cite{PRL02,wessel04}, although a vestige of the latter 
remains in the guise of local quantum criticality \cite{fermions,maxent}.

The experimental systems can be modeled by the {\em boson Hubbard} 
model \cite{Jaksch98}, described in one dimension (1D) by 
\begin{eqnarray}
\nonumber
H &=&-t\sum_{i} \left(a^{\dagger}_i a_{i+1} + a^{\dagger}_{i+1} 
a_i\right) + \mu\sum_i n_i\\ & & + V_T\sum_i x_i^2\ n_i 
+ U/2\sum_i n_i(n_i-1),
\label{hubham}
\end{eqnarray}
where $L$ is the number of sites and $x_i=ia$ is the coordinate of 
the $i$th site, and $a$ is the lattice constant. The hopping parameter, 
$t$, sets the energy scale, $n_i=a^\dagger_i a_i$ is the number 
operator, $[a_i,a^\dagger_j]=\delta_{ij}$ are bosonic creation and 
destruction operators. $V_T$ is the curvature of the trap, while 
the repulsive contact interaction is given by $U$. The chemical
potential, $\mu$, controls the number of particles. The phase diagram 
of this model in the absence of the confining trap has been extensively
studied with the goal of elucidating the various quantum phases
it exhibits \cite{FisherM,GGBRTS90,ALOTMORE} and the transitions 
between them. 

The key experimental signature of these phases lies in the interference
pattern observed after the release of the gas from the trap and subsequent
free expansion  -- an SF (MI) produces a sharp (diffuse) interference 
pattern reflecting the presence (loss) of phase coherence.
Phase coherence, especially in reduced
dimensionality, continues to be of great interest both experimentally
and theoretically. Particular attention has been focused recently on
mechanisms which can destroy quasi-long range coherence in systems on
optical lattices especially in 1D \cite{StofEss04}. 
Our focus in this Letter is the role, in 1D, of the passage from the 
SF to the MI phase in destroying phase coherence, which can
be studied in matter wave interference. 

Whereas previous studies of 
SF-MI transition focused on the height \cite{greiner} and width 
\cite{StofEss04} of the central interference peak, an alternative 
scheme was proposed recently \cite{bloch} where
the reduction of phase coherence approaching the MI was 
characterized by the visibility of interference fringes,
\begin{equation}
{\cal V} = \frac{S_{\rm max}-S_{\rm min}} {S_{\rm max}+S_{\rm min}}.
\label{vis}
\end{equation}
Here $S_{\rm max}$ and $S_{\rm min}$ are the maximum and minimum
values of momentum distribution function,
\begin{equation}
S({\bf k}) = \dfrac{1}{L}\sum_{j,l} {\rm e}^{i{\bf k}.({\bf r_j}-{\bf
r_l})}\langle a_j^{\dagger} a_l\rangle.
\end{equation}
It was observed that as the optical lattice depth [equivalent to the Hubbard
$U$ in Eq.\ (\ref{hubham})] is increased, the visibility decreases until 
special values of $U$ are reached where ${\cal V}$ displays ``kinks'' 
after which it decreases again. It was also shown \cite{bloch} that the 
values of $U$ at which such kinks are observed are reproducible, and that 
they depend on the filling (number of atoms). A perturbative treatment 
of the homogeneous MI phase \cite{bloch} has shown that ${\cal V}$ 
decreases as $U^{-1}$, improving on previous numerical results on 
small systems \cite{roth}.

Gerbier {\it et al.} \cite{bloch} proposed that the kinks are 
linked to a redistribution of the density as the SF shells 
transform into MI regions with several atoms per site. In this Letter, 
we examine in detail the presence and properties of these visibility
kinks with the help of quantum Monte Carlo (QMC) simulations of the boson
Hubbard model using the 
Stochastic Series Expansion (SSE) method \cite{sse}. We focus on 1D optical
lattices and show that while the kinks are indeed related to the
redistribution in the density associated with SF-MI transition, they
 are not solely produced by 
the transformation of SF shells into MI domains.
Indeed, we find ${\cal V}$ reveals other subtle
details of density redistributions with $U$.

\begin{figure}[b]
\begin{center}
\includegraphics[width=0.43\textwidth,height=0.43\textwidth]
{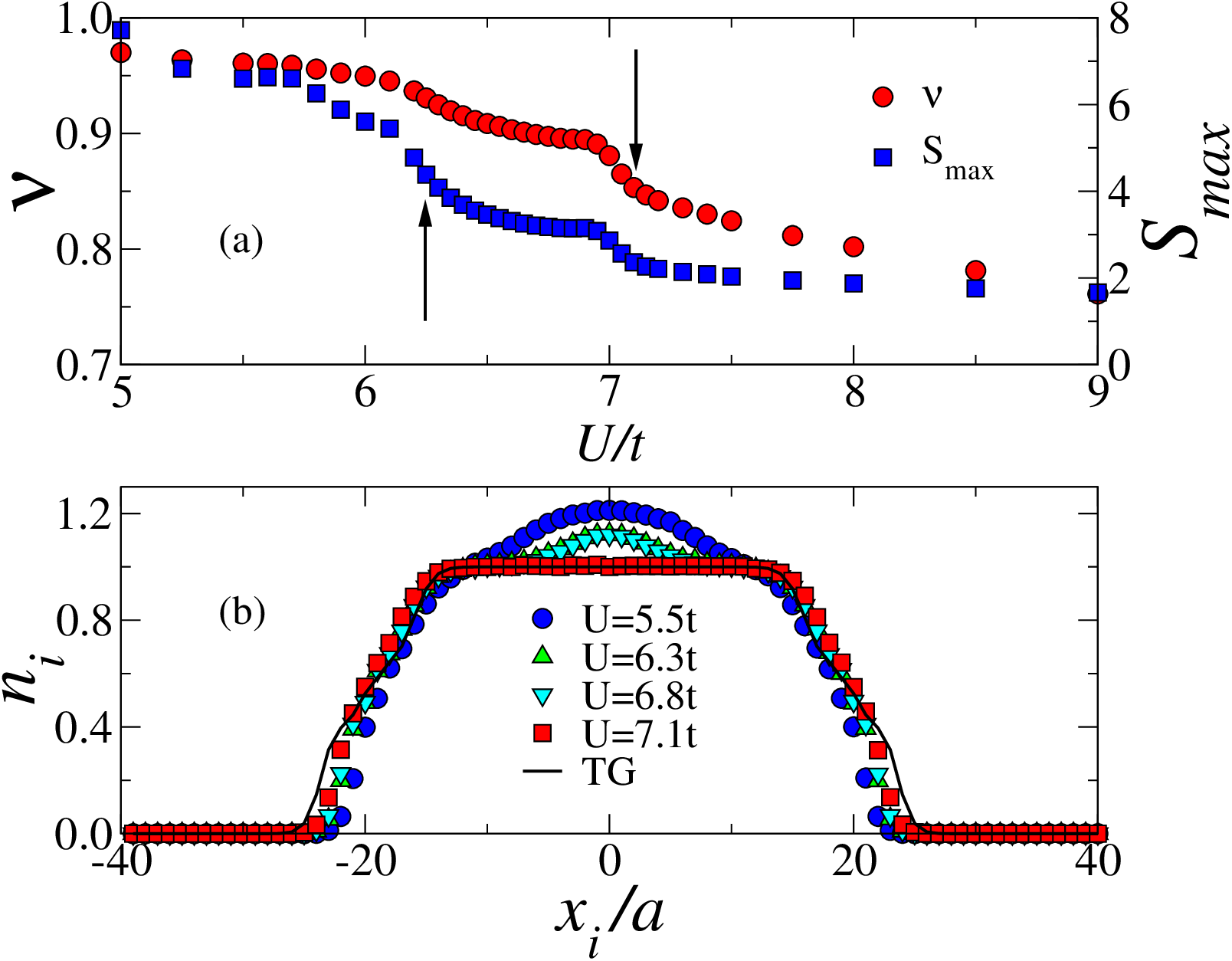}
\end{center} \vspace{-0.6cm}
\caption{(color online). (a) Visibility ${\cal V}$ and $S_{max}$ as 
functions of the on-site interaction $U/t$.  Initially, ${\cal V}$ 
decreases as $U/t$ increases. After $U/t \approx 6.3$ its rate of 
reduction decreases due to the freezing of the density profiles 
(see text). The fast decrease after $U/t \approx 7$ is related to 
the formation of the central Mott core. (b) Density profiles at 
four different values of $U/t$, and in the Tonks-Girardeau (TG) 
regime.  The profiles for $U/t=6.3$ and $6.8$
virtually coincide.  
The system under consideration has 40 bosons on a 80-site chain, and a trapping 
potential $V_T a^2 =0.01 t$.
Error bars on the data in this and all subsequent figures are smaller
than the symbol sizes.}
\label{vis1}
\end{figure}

We start our study with the simplest case, i.e., a system in which the density 
in the middle of the trap never reaches $n=2$, so that when the 
interaction strength is increased, only Mott domains with $n=1$ 
appear. In Fig.\ \ref{vis1}(a) we show the visibility 
and $S_{max}$ as functions of $U/t$. As in the experiments \cite{bloch}, 
${\cal V}$  decreases with increasing $U/t$ --  reflecting the decrease 
of $S_{max}$, and the increase of $S_{min}$ (not shown in the figure) --
with an intermediate region over which it remains fairly constant.
Two kinks can be observed both in ${\cal V}$ and $S_{max}$. The first 
one (less evident) occurs around $U/t=6.1$, and the second one 
around $U/t=7.0$.

Density profiles corresponding to four values of $U/t$ 
are depicted in Fig.\ \ref{vis1}(b).
The density profile for $U/t=6.3$ in Fig.\ \ref{vis1}(b) shows 
that the first kink in Fig.\ \ref{vis1}(a) (signaled by the first 
arrow) is related to the emergence of two MI plateaus at the 
sides ($x_i/a \approx \pm(8-12)$) of a central SF region.
The second kink in Fig.\ \ref{vis1}(a) is related to the 
formation of a full MI domain in the middle of the trap, 
which produces more evident structures in ${\cal V}$ and 
$S_{max}$. This occurs for $U/t=7.1$ as shown in Fig.\ 
\ref{vis1}(b), and signaled by the second arrow in 
Fig.\ \ref{vis1}(a). Plotting ${\cal V}$ and $S_{max}$ 
as a function of $U/t$, allows us to present more precisely the 
position and shape of the kinks: In experiments, the control 
parameter is the ratio between the lattice depth and the 
recoil energy, which produces exponential changes in 
$U/t$ \cite{bloch,Jaksch98}.

\begin{figure}[b]
\begin{center}
\includegraphics[width=0.43\textwidth,height=0.4\textwidth]
{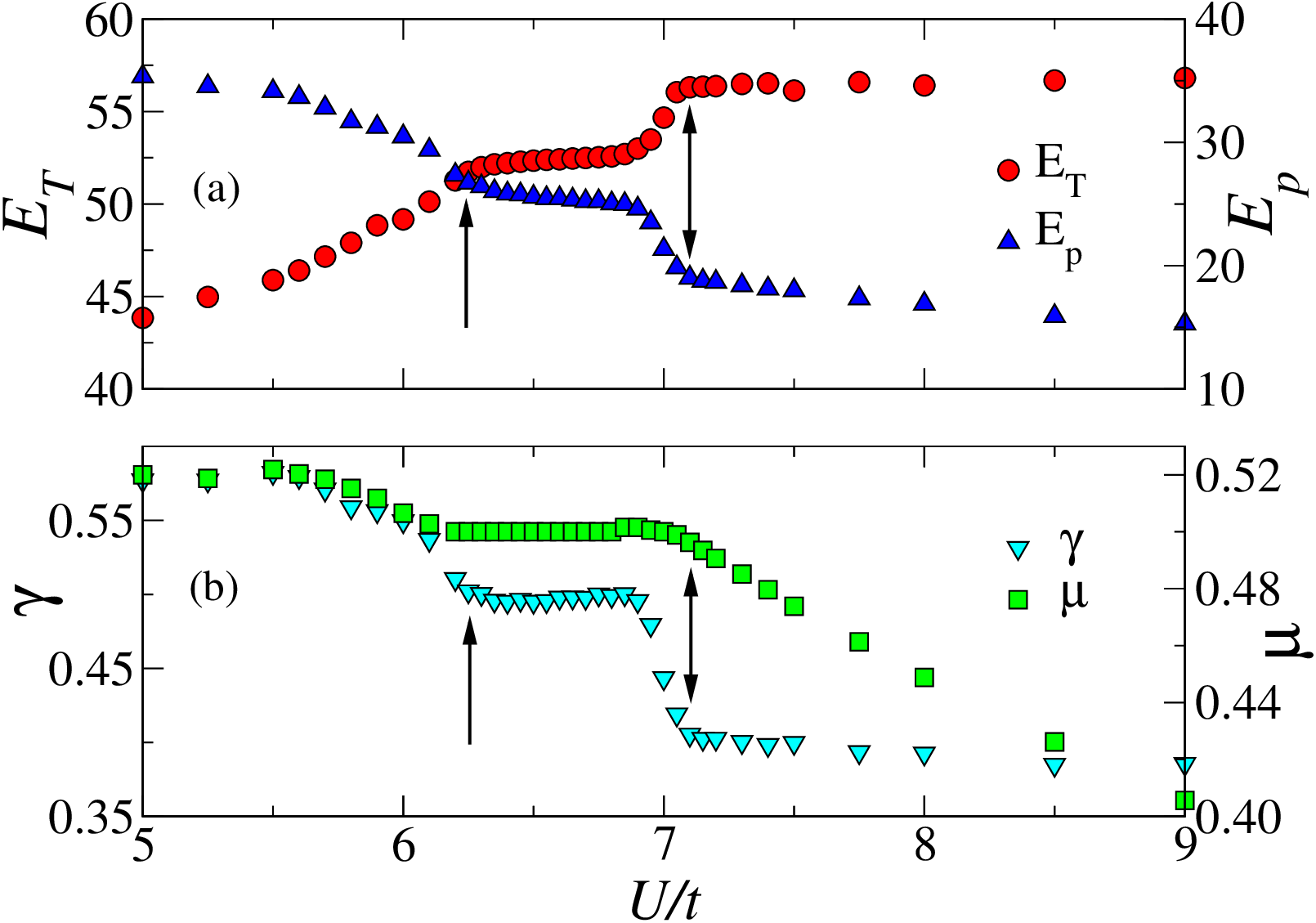}
\end{center} \vspace{-0.6cm}
\caption{(color online). (a) Trapping ($E_T$) and interaction ($E_P$) energies
as functions of $U/t$. 
(b) Ratio $\gamma= | E_p/ E_k |$ of potential to kinetic energy, and the 
chemical potential ($\mu$) needed to maintain $N_b=40$. 
The results are for the system of Fig.\ \ref{vis1}.}
\label{vis2}
\end{figure}

One unexpected feature is the freezing of the density profiles 
before the full MI forms in the middle of the trap,
which coincides with the plateau-like 
behavior of ${\cal V}$ and $S_{max}$ between the two arrows.
As $U/t$ is increased 
between 6.3 and 6.8 almost no changes occur in the density 
distribution, i.e., the bosons are no longer being pushed
out of the central regions to the outlying zones even though 
$U/t$ continues to increase. This behavior may seem 
surprising, as the central region of the system is SF, 
i.e., compressible, but can be explained by the presence of 
the emerging MI domains at the sides. The central SF region 
gets trapped between them, and the interaction $U/t$
first has to increase a finite amount before particles can be 
transferred to the SF regions at the edge against the substantially 
larger trap energy there.

This can be better understood by computing the total trapping 
($E_T$) and interaction ($E_P$) energies as functions of 
$U/t$ [Fig.\ \ref{vis2}(a)]. In the interval
$U/t=6.3$--6.8 both quantities exhibit a plateau, which is
also reflected in the chemical potential of the system 
[Fig.\ \ref{vis2}(b)]. This occurs even though the total energy 
(not shown) increases continuously, due to the 
continuous decrease in magnitude of the (negative) kinetic energy 
($E_K$) of the bosons. One can then see that the formation of 
the full MI plateau is accompanied by a fast 
increase in the total trap energy of the system by $\sim 4t$, 
the bandwidth in 1D. On the other hand, the decrease of interaction 
energy produced by the formation of the MI plateau is even 
larger $\sim 6t$. Thus, in experiments, abrupt changes can occur 
in the density profiles even if the lattice depth is increased 
slowly. This can produce the escape of particles from the trap, 
heating, or other unexpected features.

In Fig.~\ref{vis2}(b) we also show $\gamma= | E_P/ E_K |$, the ratio 
of potential to kinetic energy. This quantity is different from the one 
often used to characterize trapped bosons on lattices 
$\gamma_L=U/t$ \cite{cazalilla04}. In contrast to $\gamma_L$, for the 
system in Figs.\ \ref{vis1} and \ref{vis2}, $\gamma$ decreases with 
increasing $U$. This occurs because the density all over the trap 
becomes $n\leq 1$, and the double occupancy is strongly suppressed. 
(In the Tonks-Girardeau limit (TG), i.e., 
$U\rightarrow \infty$, $\gamma=0$ while $\gamma_L=\infty$.) Like the 
visibility and the chemical potential, $\gamma$ remains almost unchanged 
in the region where the density profiles are frozen. 

\begin{figure}[b]
\begin{center}
\includegraphics[width=0.43\textwidth,height=0.43\textwidth]
{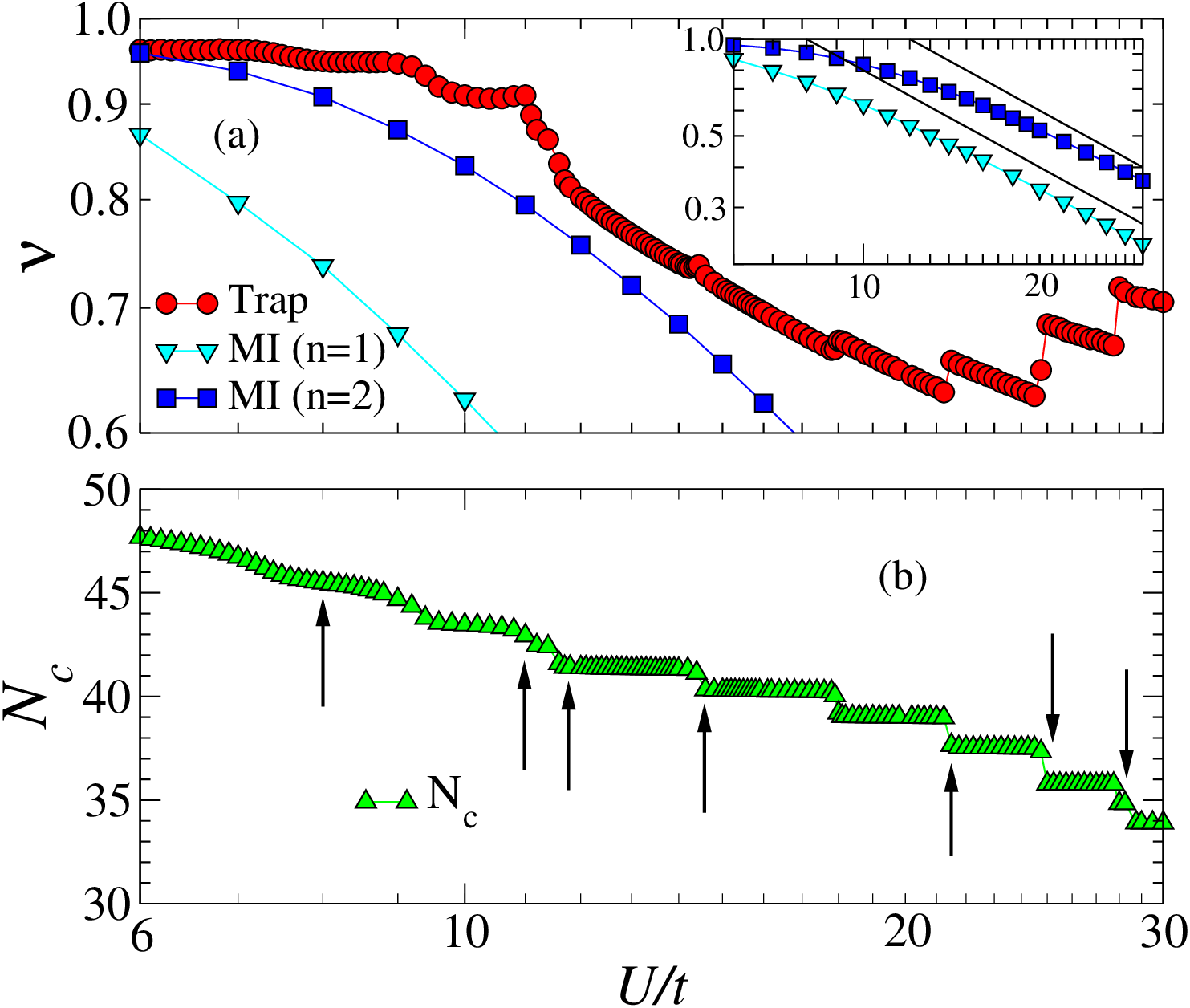}
\end{center} \vspace{-0.6cm}
\caption{(color online). (a) Visibility, ${\cal V}$, as a
function of the on-site interaction $U/t$, for $N_b=60$, and $V_T a^2 =0.06 t$, 
parameters which allow both $n=1$ and $n=2$ Mott regions to exist.
For comparison, results for
pure Mott insulating phases with $n=1$ and $n=2$ 
in open lattices without a trap are also given. 
In the inset the straight lines show the 
perturbative results of Ref.\ \cite{bloch} in 1D (see text). 
(b) Integrated density over 20 lattice sites around the center 
of the trap. }
\label{vis3}
\end{figure}

As the on-site interaction is further increased no more abrupt changes 
occur in the trap. The density profile remains almost the same, 
as seen in Fig.\ \ref{vis1}(b), where we have also plotted the exact 
result in the TG limit. The visibility and $S_{max}$, reduce continuously 
to ${\cal V}_{TG}=0.39$ and $S^{TG}_{max}=1.3$ 
(obtained using the approach presented in Ref.\ \cite{HCB}). Notice that 
even when $U\rightarrow \infty$ the visibility does not vanish, due to   
SF domains surrounding the MI.

\begin{figure}[t]
\begin{center}
\includegraphics[width=0.48\textwidth,height=0.24\textwidth]
{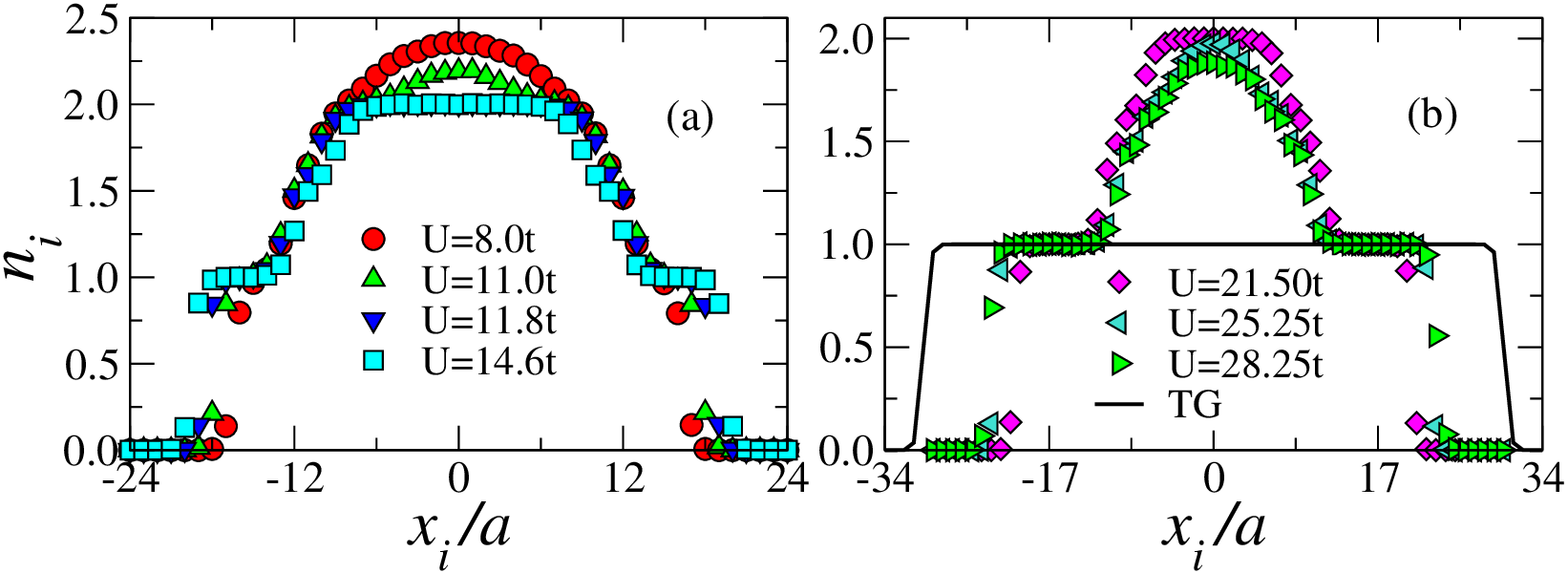}
\end{center} \vspace{-0.6cm}
\caption{(color online). Density profiles corresponding to the points
signaled by arrows in Figs.\ \ref{vis3} and \ref{vis5}. The continuous line 
in (b) is the result in the TG regime. $N_b=60$  and $V_T a^2 =0.06 t$.}
\label{vis4}
\end{figure}

When the density at the center of the trap is higher, and
exceeds two, the evolution of the visibility with
the on-site repulsion exhibits an even 
richer structure. Results for a system in that regime are 
presented in Fig.\ \ref{vis3}(a). The visibility, 
up to $U/t\sim 13$, is very similar to Fig.\ \ref{vis1}(a). 
Density profiles for three values of $U$ in that interval are presented 
in Fig.\ \ref{vis4}(a). One can see that the emergence of MI
regions with $n=1$, and $n=2$ surrounding SF regions with 
$2>n>1$, and $n>2$, respectively, produces a plateau in ${\cal V}$ due 
to a freezing of the density profiles when increasing $U$. 
In Fig.\ \ref{vis3}(a), the formation of the $n=2$ plateau abruptly 
reduces the visibility similar to the formation of the $n=1$ plateau in 
Fig.\ \ref{vis1}(a).

However, the behavior above $U/t=13$ has additional structures
compared to Fig.\ \ref{vis1}(a). 
In order to understand the origin of these visibility features,
we have plotted in Fig.\ \ref{vis3}(b) the 
integrated density over 20 lattice sites around the center of the 
trap $N_c=\sum_{i=-10}^{10} n_i$. A clear one to one mapping 
between the features in the visibility and plateau in $N_c$ is seen. 
The visibility kinks result not from the 
formation of new SF or MI regions, but rather 
from a redistribution of bosons between the MI states with 
$n=2$, and $n=1$. As seen in Figs.\ \ref{vis3}(b), and \ref{vis4}, 
such a redistribution occurs discontinuously in 
$U$. In addition, since the SF domains with $2>n>1$ can 
increase their sizes during such a process, 
the visibility can increase [see for example the kinks around 
$U/t=14.6$ and 21.5 in Fig.\ \ref{vis3}(a), and the 
corresponding density profiles in Fig.\ \ref{vis4}].

The above features are not restricted to the 1D character of the system,
and could be observed in higher dimensions. However, as $U$ is 
increased even further $(U\gtrsim 25t)$, a purely 1D effect sets in. As the MI
plateau with $n=2$ melts, correlations start to develop between 
the two disconnected SF domains with $2>n>1$. This produces
a large increase of the visibility, as seen in Fig.\ \ref{vis3}(a).
[The corresponding density profiles are shown in Fig.\ \ref{vis4}(b).]
In 1D this increase in the visibility provides an 
unambiguous signature of the presence, and melting, of the $n=2$
(or larger) MI domain. This can be useful for understanding
the dynamics of strongly correlated bosons in 1D \cite{damping}.

For very large values of $U$, beyond the ones in Fig.\ \ref{vis3}(a), 
the $2>n>1$ SF domain will eventually disappear, as occurs in 
Fig.\ \ref{vis1}(a), producing a further reduction in the visibility. 
In the TG regime, we obtain (for these parameters) ${\cal V}_{TG}=0.02$. 
The corresponding density 
profile can be also seen in Fig.\ \ref{vis4}(b).

We have also plotted in Fig.\ \ref{vis3}(a), the values obtained for the 
visibility in homogeneous systems with 60 bosons and densities $n=1$
and $n=2$. (We have used open boundary conditions as they are the closer
to the trapped case). These results in homogeneous systems are very 
different from the ones in the trapped case. Due to the existence 
of SF domains, the visibility in the trap is always larger than that in 
the homogeneous case. In the region of interest, where
the MI plateau emerges, and melts, no extrapolation is possible from 
the uniform case.  Only for very large values of $U$, after a MI domain 
appears in the center of the trap, can one can expect the uniform
and trapped systems to behave similarly. In the inset we have compared 
the results for the homogeneous systems, with those obtained
in Ref.\ \cite{bloch} 
[${\cal V}_{1D}=4(n+1)t/U$]. For the largest values of $U$ one can see that
the $t/U$ power law starts to develop, but its prefactor is still 
different from $4(n+1)$, so that very large values of $U$ are needed
for a good agreement in 1D.

\begin{figure}[t]
\begin{center}
\includegraphics[width=0.42\textwidth,height=0.23\textwidth]
{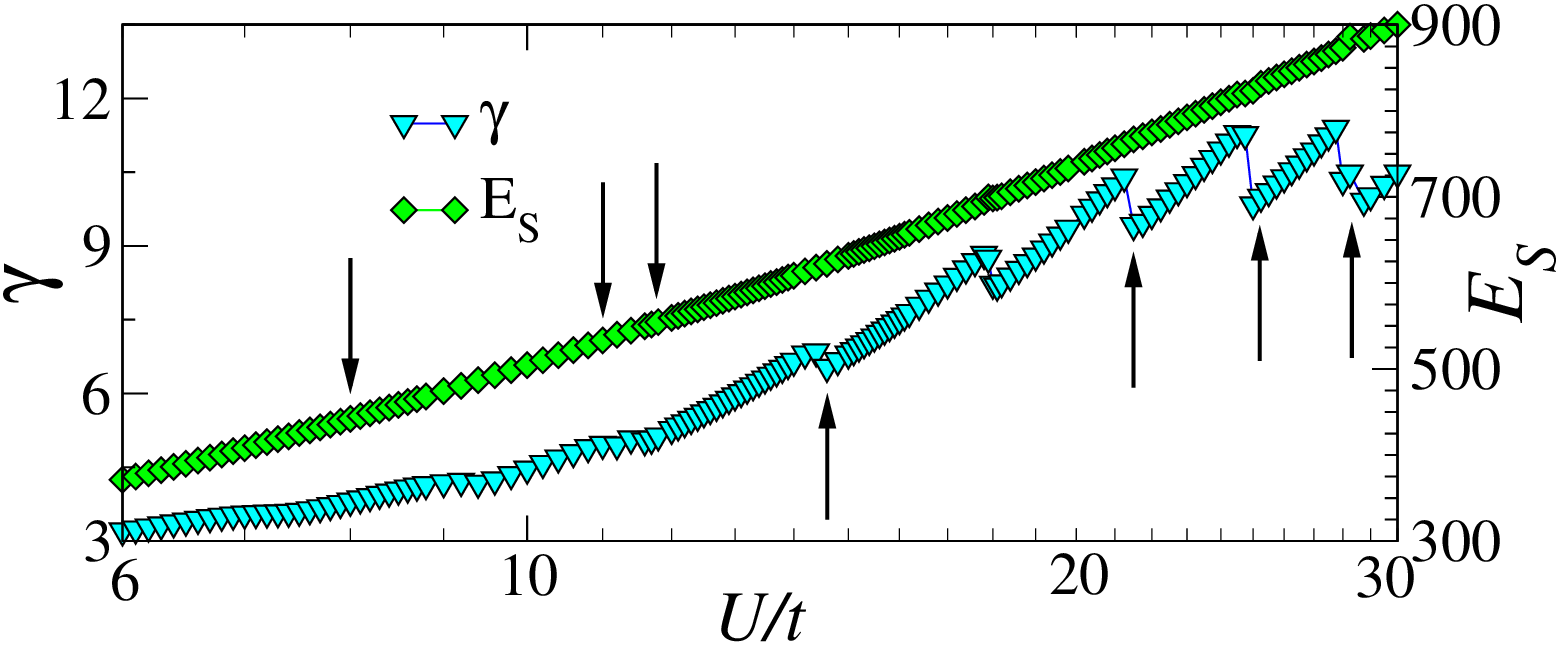}
\end{center} \vspace{-0.6cm}
\caption{(color online). Ratio $\gamma= | E_p/ E_k |$ 
of potential to kinetic energy, and the total energy of the system
($E_S$) vs $U/t$, for the system of Figs.\ \ref{vis3} and \ref{vis4}.
$N_b=60$ and  $V_T a^2 =0.06 t$.}
\label{vis5}
\end{figure}

We conclude by showing in Fig.\ \ref{vis5} the behavior of 
$\gamma$ in the system of Figs.\ \ref{vis3} and \ref{vis4}.
In this case, since the density at the middle of the trap is 
larger than one, i.e., there is significant double occupancy in this 
region, $\gamma$ (and $E_P$) increases with $U/t$. It also exhibits 
the same jumps produced by the redistribution of particles in 
Fig.\ \ref{vis4}. As in the system in Figs.\ \ref{vis1} and 
\ref{vis2}, this occurs even when the total energy of the system 
($E_S$) increases continuously with $U$, as can be also seen in 
Fig.\ \ref{vis5}.

In this Letter we have explored the evolution of the visibility of 
trapped atomic gases in one dimensional optical lattices using
Quantum Monte Carlo simulations. We have shown that the visibility 
behaves very similar to that observed experimentally. In particular, 
it has kinks associated with redistribution of density amongst Mott 
insulating and superfluid regions within the trap. In addition, we have 
also exhibited several other novel features of the visibility evolution 
in 1D, like a large increase due to the melting, with increasing $U/t$, of  
$n>1$ MI plateaus. We have demonstrated that the evolution of the density 
distribution with interaction strength exhibits pauses. That is, at 
certain values of $U$ the density distribution, and other observables, 
do not change even when the interaction strength increases over a 
range as large as $t/2$. We have shown that the emergence of this 
static behavior is associated with the formation of Mott insulating 
plateaus away from the trap center. These plateaus block the transfer 
of bosons to the outer parts of the system, and hence cause 
the evolution to stall. While many quantities in trapped Bose systems 
are well described by the local density approximation, it is not clear 
that approach will capture the above behavior, including the kinks in 
the visibility. This is because these effects are intrinsically tied to the 
competition of the trap versus kinetic and interaction energies in 
systems where the SF and MI domains are of finite width, as in the ones 
explored in the recent experiments.
 
We thank I. Bloch and M. Troyer 
for helpful discussions. M.R. and R.T.S. were 
supported by NSF-DMR-0312261. P.J.H.D. by Stichting FOM.

\end{document}